\documentclass[pra,superscriptaddress,longbibliography,twocolumn]{revtex4-1}
\usepackage[utf8]{inputenc}
\usepackage[english]{babel}
\usepackage{amsmath}
\usepackage{amsfonts}
\usepackage{amssymb}
\usepackage{graphicx}
\usepackage{amsthm}
\usepackage{color}
\usepackage{hyperref}
\usepackage[caption=false]{subfig}
\usepackage{graphicx,float}% Include figure files
\usepackage{dcolumn}% Align table columns on decimal point
\usepackage{bm}% bold math
\usepackage{mathrsfs}
\usepackage{txfonts}
\usepackage{CJK}
\usepackage[amssymb]{SIunits}
\usepackage{epsfig}

\usepackage{epstopdf}
\usepackage{lipsum}
\usepackage{placeins}
\usepackage{physics}

\newenvironment{SChinese}{%
\CJKfamily{gbsn}%
\CJKtilde
\CJKnospace}{}
\allowdisplaybreaks[2]

\begin{document}

\begin{CJK}{UTF8}{}
\begin{SChinese}

\title{A table-top high-sensitivity gyroscope based on slow light and cavity enhanced photon drag}
%\title{Bypassing the purity-yield trade-off in Heralded Single-Photon Sources via Photon Blockade}

\author{Min She}  %
 \affiliation{College of Engineering and Applied Sciences, National Laboratory of Solid State Microstructures, Nanjing University, Nanjing 210023, China}

\author{Jiangshan Tang}  %
 \affiliation{College of Engineering and Applied Sciences, National Laboratory of Solid State Microstructures, Nanjing University, Nanjing 210023, China}

\author{Keyu Xia}  %
  \email{keyu.xia@nju.edu.cn}
  \affiliation{College of Engineering and Applied Sciences, National Laboratory of Solid State Microstructures, Nanjing University, Nanjing 210023, China}

%\date{\today}

\begin{abstract}
A high-sensitivity gyroscope is vital for both investigation of the fundamental physics and monitor of the subtle variation of Earth's behaviors. However, it is challenge to realize a portable gyroscope with sensitivity approaching a small fraction of the Earth's rotation rate. Here, we theoretically propose a method for implementing a table-top gyroscope with remarkably high sensitivity based on photon drag in a rotating dielectric object. By inserting an $\text{Er}^{3+}$-doped glass rod in a Fabry-P\'{e}rot optical cavity with only $20~\centi\meter$ length,  we theoretically show that the giant group refractive index and the narrowing cavity linewidth due to slow light can essentially increase the nonreciprocal phase shift due to the photon drag to achieve a rotation sensitivity of $26~\femto\rad /\second/\sqrt{\hertz}$. This work paves the way to accurately detect tiny variations of the Earth's rotation rate and orientation, and even can test the geodetic and frame-dragging effects predicted by the general relativity with a small-volume equipment.
\end{abstract}

\maketitle

\end{SChinese}
\end{CJK}

%Introduction
\section{Introduction}
The sensitive gyroscope measuring rotation of an object or the frame can be used to test fundamental physical effects~\cite{1.Gen Relat Gravit 20,2.Phys. Rev. Res. 2} and also promise important applications such as directional positioning~\cite{3.J Microelectromech Syst 22,4.Sensors (Basel) 23}, inertial navigation systems~\cite{5.Sensors (Basel) 19,6.Opt Express 30}, and sensing length-of-day variation of the Earth~\cite{7.Opt Express 28, 8.Nat. Photonics 14}. Currently, various methods are developed to measure slow rotation by using microelectromechanical gyroscopes~\cite{9.Meas Sci Technol 34, 10.J. Microelectromech. Syst. 15, 11.J. Microelectromech. Syst. 27,12.Sens. Actuator A Phys. 224,13.IEEE Trans. Ind. Electron. 65}, surface acoustic wave gyroscopes~\cite{14.Sensors (Basel) 15, 15.Opt. Express 26, 16.Appl. Phys. Express 4,17.Microsyst. Nanoeng. 8}, ring laser gyroscopes (RLGs)~\cite{18.Phys. Rev. Lett. 125, 19.Nature 576, 20.Opt. Lett. 44, 21.Nat. Commun. 11}, fiber optic gyroscopes (FOGs)~\cite{22.Opt. Lett. 48, 23.Opt. Express 27, 24.Opt. Lett. 46}, atomic-optical hybrid gyroscopes~\cite{25.Phys. Rev. Lett. 124, 26.Phys. Rev. Appl. 14}, and nanophotonic optical gyroscopes~\cite{27.Nat. Photonics. 12}. Among them, optical gyroscopes based on the Sagnac effect are most reliable and show the highest sensitivity thus far. Its principle involves two counter-propagating waves in the same closed loop, generating a beat frequency reflecting the rotation rate~\cite{28.Seances Acad. Sci. 157}.  

A Sagnac-effect-based gyroscope can reach a high sensitivity but typically requires a large size. The ROMY RLG with six $12~\meter$ arms displays a sensitivity of $\sim 80~ \pico \rad/ \second/ \sqrt {\hertz}$~\cite{ 29.Geophys J. Int. 225,30.Phys. Rev. Lett. 125}. Thus, it can analyze small ground disturbances caused by earthquakes and oceans. The GINGERINO with four $3.6~\meter$ long sides shows a sensitivity up to the level of $\pico \rad/ \second/ \sqrt {\hertz}$~\cite{31.arxiv-2301.01386}. The ring gyroscope G using a $16$-\meter-circumference ring cavity achieves $12~\pico \rad/ \second/ \sqrt {\hertz}$ and a resolution of several milliseconds in detecting disturbances on the Earth's surface after $120$ days of continuous measurements~\cite{32.Nat. Photonics 17}. These RLGs utilize mirrors with ultra-low optical loss to construct high finesse optical cavity and use two counter-propagating light beams to reduce bias drift errors~\cite{33.J. Appl. Phys. 105, 34.Eur. Phys. J. C 82}. Despite a great success in rotation measurement, these RLGs are large in size thus far, costly and fixed on a basis. In contrast, FOGs shrink the volume by circling a long optical fiber to many-turn coil and therefore are portable, but the obtained sensitivity is relatively lower. The state-of-the-art FOG can only achieve a sensitivity of $9.7 \times 10^{-8}~\rad/ \second/ \sqrt {\hertz}$~\cite{35.Opt. Lett. 48}, several orders lower than the RLG counterparts. Because of the long optical path, the application of FOGs is constraint by the asymmetric excitation in the middle of the fiber circuit caused by temporal and local perturbation, resulting in the non-negligible dynamic errors~\cite{36.Opt. Fiber Technol. 69,37.Phys. Usp. 45}. Such asymmetric perturbation is a common challenge in the Sagnac-effect-based gyroscope with a big ring. Precise measurement of daily variations of the Earth's rotation rate ($\Omega_\text{e} = 72.92~\micro\rad/\second/\sqrt{\hertz}$) requires a sensitivity approaching the level of tens $\pico\rad/\second\sqrt{\hertz}$, which has only been achieved by the large-size RLG~\cite{2.Phys. Rev. Res. 2, 30.Phys. Rev. Lett. 125, 31.arxiv-2301.01386, 32.Nat. Photonics 17}. Therefore, it is highly desirable to develop a portable gyroscope with such sensitivity but remains a challenge~\cite{38.Science. 356}.

Propagation of linearly-polarized light in a spinning medium results in the change of polarization plan, known as the photon drag effect, because the rotation of medium causes circular birefringence~\cite{39.Proc. R. Soc. London Ser. A 349.441}. The slow-light medium can significantly enhance the photon drag and narrow the linewidth of a cavity by several orders~\cite{40.Science 333, 41.Phys. Rev. Lett. 111}. However, it is thus far yet to explore the interplay of slow light and the photon drag effect for a sensitive gyroscope. 

In this work, we propose a photon-drag-based gyroscope by embedding a slow-light medium in an $20~\centi\meter$ long FP cavity. By measuring the transmitted light power of the cavity, we can, in principle, achieve the remarkable gyro sensitivity. Our proposal provides a table-top device for measuring daily variations in the Earth's rotation rate and testing the effects of Lorentz Violation and General Relativity on earth.

The paper is organized as follows: The system and model are described and developed in section II. Then, the results are presented in section III. The feasibility of implementing our system is discussed in the section ``Implementation''. In the end, the paper is concluded in section V.

\section{System and model}
% Model system 1
\begin{figure}
	\centering
	\includegraphics[width=1.0\linewidth]{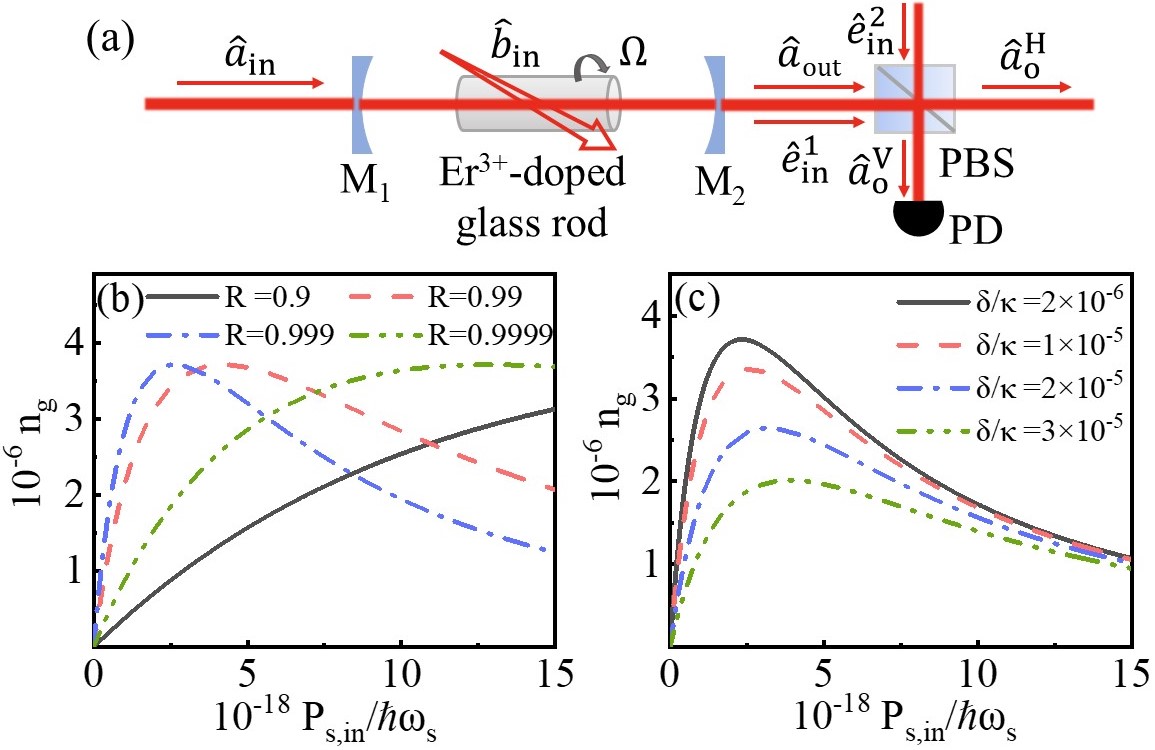} \\
	\caption{(a) Schematic of gyroscope based on the slow-light enhanced photon drag in a high-Q FP cavity embedded with an $\text{Er}^{3+}$-doped glass rod spinning at a rate $\Omega$. A HP light is input to the cavity. The VP transmitted laser field is detected by a photodetector (PD). (b) Group refractive index versus the input photon flux under different mirror reflectivity when $\delta = 10~\hertz$. (c) Group refractive index with input photon flux at various frequency detuning when the mirror reflectivity $R=0.998$ and $\kappa = 5$~\mega\hertz. Other parameters: $L=10~\centi\meter$ and $L_0=20~\centi\meter$, yielding $\eta=0.5$.}
	\label{fig:FIG1}
\end{figure}

The system of our gyroscope is depicted in Fig.~\ref{fig:FIG1}(a). An $\text{Er}^{3+}$-doped glass rod is inserted into the FP cavity form by two highly reflective mirrors $\text{M}_1$ and $\text{M}_2$. By detecting the field component orthogonal in polarization to the input field, this system can work as a gyroscope.  
We assume that the two mirrors have the same reflectivity $R$ and cause a decay rate $\kappa_\text{e}$ to the cavity each. We also assume that the cavity has an intrinsic loss with rate $\kappa_\text{i}$. The overall decay rate of the cavity is then given by $\kappa = 2\kappa_\text{e} + \kappa_\text{i}$.
The two ends of the rod is coated with anti-reflective film. A horizontally-polarized (HP) probe laser beam, denoted as $\hat{a}_\text{in}$, is input to the cavity via the $\text{M}_1$ mirror. $\hat{a}_\text{out}$ denotes the transmitted field. When the frame or the glass rod spins at a rate $\Omega$, the photon drag causes difference $\Delta n$ between the refractive indices $n_\text{l}$ and $n_\text{r}$ ``seen'' by the left-hand circularly polarized (LCP) and the right-hand circularly polarized (RCP) cavity eigenmodes, respectively. Because of the unidirectional mechanical angular momentum, this difference $\Delta n$ leads to nonreciprocal phase shift and lifts the resonance frequencies of the LCP and RCP eigenmodes to non-degenerate. As a result, the LCP and RCP components of the input field accumulate a non-zero relative phase after transmitting through the cavity. A vertically-polarized (VP) component $\hat{a}^V_\text{o}$ appears in the transmitted field, denoted as $\hat{a}_\text{out}$. It is then separated by a polarization beam splitter (PBS). By detecting its power with a photodetector, we can determine the spinning rate $\Omega$. Besides, quantum noise ($\hat{e}_1$ and $\hat{e}_2$) also enters the photodetector and causes error.

We consider the slowly rotating frame, e.g. Earth's rotation. Then, the photon drag is very weak. To amplify the signal, another control laser beam $\hat{b}_\text{in}$ is used to increase the group velocity index of the probe field to $n_\text{g} \gg 1$ in the glass rod via the coherent population oscillation (CPO) process of the $\text{Er}^{3+}$ ions.  The enhancement resulting from slow light is twofolds: enhancing the photon drag by several orders~\cite{40.Science 333} and significantly narrowing the cavity linewidth~\cite{41.Phys. Rev. Lett. 111}. The combination of both allows us to make a table-top gyroscope with sensitivity exceeding  the large RLGs. Here, we neglected the influence of the photon drag due to the spin of the mirrors and the PBS, because it is much smaller than the slow-light enhanced photon drag in the $\text{Er}^{3+}$-doped glass rod.

Now we discuss how to enhance the gyroscope sensitivity by using slow light in a FP cavity. In the absence of rotation, the control and probe laser beams act  on the $\text{Er}^{3+}$-doped glass rod together. At this time, $\text{Er}^{3+}$ ions can be approximated as a two-level system with transition ${}^4I_{15/2}\to {}^4 I_{13/2}$ at $\lambda  \approx 1536~\nano\meter$~\cite{42.J. Opt. Soc. Am. B 28,43.Opt. Commun 279}. The transition frequency is $\omega$. $T_1$ is the relaxation time of the excited state. The burnt hole can appear in the probe spectrum via the CPO process. It can cause a giant group refractive index $n_\text{g}$ given by~\cite{44.Phys. Rev. Lett. 90,45. Phys. Rev. A 24} (see the Appendix A).
\begin{equation}\label{eq:ng}
	n_\text{g}=n_0+\frac{\alpha_0cT_1}{2}\frac{I_0}{1+I_0}\left[\frac{1}{\left(T_1\delta\right)^2+\left(1+I_0\right)^2}\right],
\end{equation}
where $n_0$ is the refractive index of the host medium, $\alpha_0$ is the unsaturated absorption coefficient, $\delta=\omega_\text{p}-\omega_\text{s}$ with $\omega_\text{p}$ and $\omega_\text{s}$ being the frequencies of the probe and control fields, respectively. $I_\text{sat}$ is the saturated absorption intensity of the ions. $I_\text{p}$ and $I_\text{s}$ denote the intensity of the probe and control field, respectively. $n_\text{g}$ is crucially dependent on the ratio $I_0=I_\text{s}/I_\text{sat}$. The saturated absorption intensity of the ions is given by
\begin{equation}\label{eq:Isat}
	I_\text{sat}=\hbar\omega/[T_1 (\sigma_\text{12}+\sigma_\text{21})] \;.
\end{equation}
The operators $\sigma_{12}$ and $\sigma_{21}$ are the absorption cross-section and emission cross-section, respectively, $\sigma_{12}$ $\approx$ $\sigma_{21}$. The intensities of the control fields in the cavity is given by 
\begin{equation} \label{eq:Is}
	I_\text{s} =c\varepsilon_0E_0^2 |b_\text{in}|^2 \kappa_{e}/\kappa^2 \;, 
\end{equation} 
where $E_0=\sqrt{\hbar \omega_\text{s}/\varepsilon_0 V}$, $\varepsilon_0$ is the dielectric constant, $V$ is the mode volume. The input powers of the control and probe fields are $P_\text{s,in} = \hbar \omega_\text{s} |b_\text{in}|^2$ and $P_\text{p, in} = \hbar \omega_\text{p} |a_\text{in}|^2$, respectively. Throughout the following investigation, we fix the power ratio to $P_\text{p, in}/ P_\text{s, in} = I_\text{p}/ I_\text{s} = 0.08$, which is accessible in experiment~\cite{46.Europhys. Lett. 73}. In experiment, one can modulate the input laser beam to form the control and probe fields. The power or intensity ratio between the control beam and the probe beam can be tuned via modulation. In our system, the modulation frequency is very small such that $\omega_\text{p} \approx \omega_\text{s}$ to guarantee a giant group refractive index $n_\text{g}$.  For an $\text{Er}^{3+}$-doped glass rob, we have $n_0 = 1.48$, $T_1=10.5~\milli\second$ and $\alpha_0=0.16~\centi\meter^{-1}$~\cite{43.Opt. Commun 279}.

Figure~\ref{fig:FIG1}(b) reflects the group refractive index $n_\text{g}$ as a function of the power of the control laser field for various mirror reflectivity. The peak values of $n_\text{g}$ are very close for different reflectivities but the required control laser power, $P_\text{s, in}$, increases as the reflectivity reduces. When $R=0.99$, the peaked $n_\text{g} \approx 3.7~\times 10^6$ is attained at $P_\text{s, in} = 440~\milli\watt$. When the reflectivity is improved to $R = 0.999$, which is much harder, the peaked $n_\text{g}$ is about $3.7~\times 10^6$ and the optimal power $P_\text{s, in}$ decreases to $\sim 272~\milli\watt$, corresponding to $2 \times 10^{18}$.  Thus, it is experimentally convenient to choose the reflectivity between $0.99 - 0.999$.  Fig.~\ref{fig:FIG1}(c) shows the dependence of $n_\text{g}$ on the input photon flux of the control light for various frequency difference $\delta$. The peak $n_\text{g}$ first increases rapidly and then decreases after reaching the maximum value as the control laser power increases. The peak value also decreases as the modulation frequency $\delta$ increasing. From Figs.~\ref{fig:FIG1}(b) and (c), the group refractive index can reach to the level of $10^6$.

We learn from Ref.~\cite{41.Phys. Rev. Lett. 111} that slow light can narrow the linewidth of the cavity. The total loss can be approximated as $\kappa'\approx\kappa/n_\text{g}$ and $\kappa_\text{e}'\approx\kappa_\text{e}/n_\text{g}$. According to the input-output relationship of the optical cavity~\cite{47.Phys. Rev. A 31,48.Phys. Rev. X 3}, the probe filed output operator can be expressed as
\begin{equation} \label{eq:aout}
	\hat{a}_\text{out}=\frac{2\hat{a}_\text{in}\kappa_\text{e}}{i\Delta/n_\text{g}+\kappa} \;,
\end{equation}
where $\Delta=\omega_\text{c}-\omega_\text{p}$, $\omega_\text{c}$ is the cavity frequency. In the case of rotation, the rotation of the polarization state is caused by the refractive index difference between LCP and RCP light~\cite{39.Proc. R. Soc. London Ser. A 349.441}. The refractive index difference between LCP and RCP light is given as
\begin{equation}\label{eq:Delta_n}
	\Delta n=n_\text{r}-n_\text{l}=\left(n_\text{g}-\frac{1}{n_{\phi}}\right)\frac{\Omega}{\omega_\text{c}},
\end{equation}
where $n_\phi$ is the phase refractive index, $\Omega$ is the medium rotation rate, $n_\text{r}$ and $n_\text{l}$ are the refractive indices for RCP and LCP light, respectively. Based on the resonance condition of the cavity, the phase change can be expressed as
\begin{equation}\label{eq:Delta_omega}
	\Delta\omega_\pm=\mp\left(n_g-\frac{1}{n_\phi}\right)\Omega\eta,
\end{equation}
where $\eta=L/L_0$ represents the duty ratio, $L_0$ and $L$ are the cavity length and medium length, respectively. At this point, the probe filed output operator can be approximated as 
\begin{equation}\label{a_out}
	\hat{a}_\text{out}^\pm=\xi\hat{a}_\text{in} e^{-i\Delta\omega_\pm n_\text{g}/ \kappa} \;,
\end{equation}
where $\xi=2\kappa_\text{e}/\kappa$. Subsequently, the output light from the FP cavity is split into the HP and VP components after passing through the PBS. Using the relationship between circularly and linearly polarized light, we obtain the output of VP component is
\begin{equation}\label{eq:a_o}
	\hat{a}_\text{o}^V=-\xi\hat{a}_\text{in}\sin\left(\Delta\phi\right)+\hat{e}_\text{in},
\end{equation}	
where $\Delta\phi=\left(\Delta\omega_+-\Delta\omega_-\right)/2\kappa'$, $\hat{e}_\text{in}=\left(\hat{e}_\text{in}^1+\hat{e}_\text{in}^2\right)/\sqrt{2}$ is the quantum fluctuation. We define the HP component output photon number operator as
\begin{equation}\label{eq:m_o}
	\hat{M_\text{o}}=\xi^2\sin^2(\Delta\phi)\hat{a}_\text{in}^\dagger \hat{a}_\text{in}+\hat{e}_\text{in}^\dagger \hat{e}_\text{in}-\xi\sin(\Delta\phi)(\hat{a}_\text{in}^\dagger \hat{e}_\text{in}+\hat{e}_\text{in}^\dagger \hat{a}_\text{in}) \;.
\end{equation}

Based on the error propagation analysis~\cite{26.Phys. Rev. Appl. 14,49.Opt. Express 24}, the rotation sensitivity is defined as 
\begin{equation}\label{eq:Delta_Omega1}
	\Delta\Omega={\langle\Delta M_\text{o}^2\rangle^{1/2}}/{|\partial\langle \hat{M_\text{o}} \rangle/\partial\Omega|} \;,
\end{equation}
where 
\begin{subequations}
	\begin{align}
		\langle\Delta M_\text{o}^2\rangle & =\langle \hat{M_\text{o}^2}\rangle-\langle \hat{M_\text{o}}\rangle^2 \approx \xi^2\sin^2(\delta\phi)N_\text{in} \;, \\
		\left|{\partial\langle \hat{M_\text{o}}\rangle}/{\partial\Omega}\right| & =2\xi^2 |\sin(\Delta\phi)\cos(\Delta\phi)| n_\text{g}^2 N_\text{in}\eta/{\kappa},
	\end{align}
\end{subequations}
with  $N_\text{in}=|a_\text{in} |^2$ being the input probe photon flux. According to the above formula, our system sensitivity $\Delta\Omega$ is given by
\begin{equation}\label{eq:Delta_Omega2}
	\Delta\Omega=\frac{\kappa}{2\xi\eta n_\text{g}^2\sqrt{N_\text{in}}}.
\end{equation}

The gyro sensitivity relies on the input photons flux $(N_\text{in})$, the group refractive index $(n_g)$, the duty ratio $(\eta)$, and the decay rate $(\kappa)$. In order to obtain optimal sensitivity, the optimal operation is discussed next.

\begin{figure}
	\centering
	\includegraphics[width=1.0\linewidth]{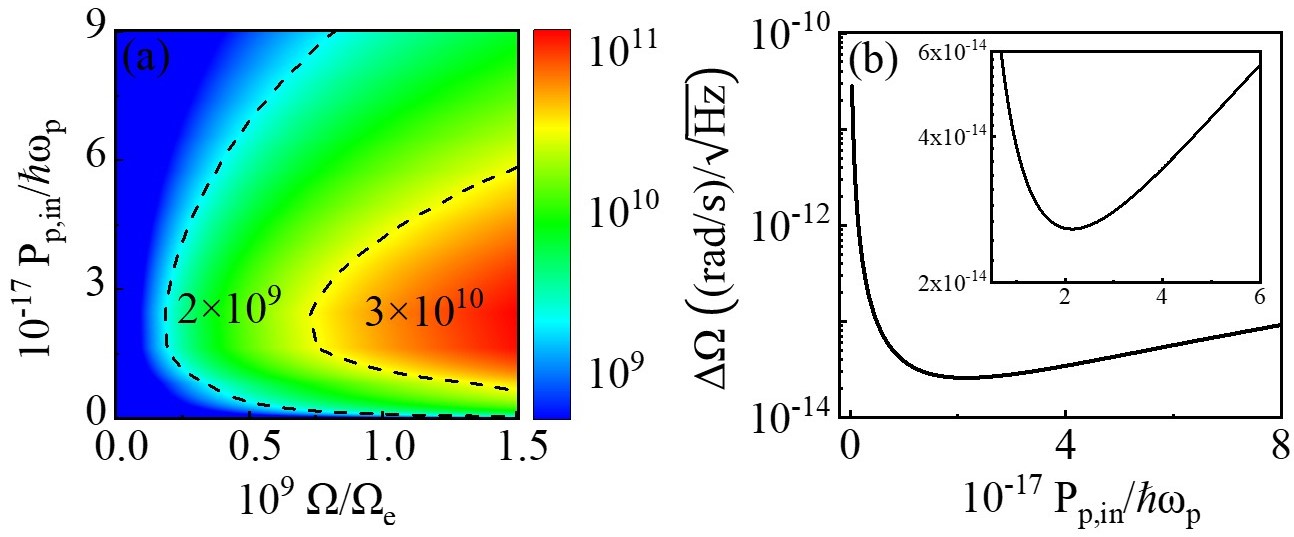} \\
	\caption{(a) Photon flux (color bar) of the VP transmitted light as a function of the rotation rate of the rod and the input photon flux. (b) Sensitivity of the gyroscope as a function of the input photon flux. Other parameters as in Fig.~\ref{fig:FIG1} except for $R=0.998$.}
	\label{fig:FIG2}
\end{figure}

\section{Results}
The performance of our gyro crucially relies on the power of input probe light. Figure~\ref{fig:FIG2} shows the photon flux of the output VP component and the gyroscope sensitivity. According to the photon drag effect, the deflection phase caused by a rotating medium is proportional to the group refractive index $n_\text{g}$~\cite{50.Phys. Rev. Lett. 100}. Thus, the giant group refractive index induced in the $\text{Er}^{3+}$-doped glass rod can essentially enhance the polarization rotation of light due to the photon drag~\cite{40.Science 333}, greatly amplifying the output power to a detectable level. Figure~\ref{fig:FIG2}(a) shows the output photon flux as a function of the probe power and the rotation rate. Taking the experimentally accessible values for parameters $R=0.998$, $L=10~\centi\meter$, $L_0=20~\centi\meter$, $\eta=0.5$, the maximum output photon flux increases with the rotation rate. It reaches $P_\text{out}/\hbar\omega_\text{p} \approx 10^{11}~\text{photon/s}~(\approx 17~\text{nW})$ when $\Omega/\Omega_\text{e}\approx10^{-9}$. This level of photon flux can be well detected experimentally. The peak output photon flux is attainable for $\Omega/\Omega_\text{e}<1.5 \times 10^{-9}$. The required input power less than $P_\text{p,in}/\hbar\omega_\text{p} <4 \times 10^{17} \text{photons}\per\second$, corresponding to an probe field power $< 50 ~\milli\watt$. The performance of our gyroscope is determined by the duty ratio $\eta$ but not the cavity length $L_0$. Throughout the following investigation, we fix $L_0 = 20~\centi\meter$. 

According to Eq.~(\ref{eq:Delta_Omega2}), it can be seen that the gyro sensitivity is proportional to $\kappa$ and inversely proportional to $\eta$, $n_\text{g}^2$, and $\sqrt{N_\text{in}}$. Therefore, the sensitivity tends to be improved as the input power increases, as shown in Fig.~\ref{fig:FIG2}(b), but reaches the optimal value of $\Delta\Omega_\text{min}=26~\femto\rad /\second/\sqrt{\hertz}$ when the input photon flux $P_\text{p,in}/\hbar\omega_\text{p} \approx 2 \times 10^{17}~\text{photons\per\second}~(\approx 27~\milli\watt)$, and then reduces. The reason is that $n_\text{g}$ first increases rapidly and then decreases after reaching the maximum value as the input power increases, see Figs.~\ref{fig:FIG1}(b) and (c). It implies that, to obtain the best sensitivity, the gyro needs to operate at the optimal input power.

\begin{figure}
	\centering
	\includegraphics[width=1.0\linewidth]{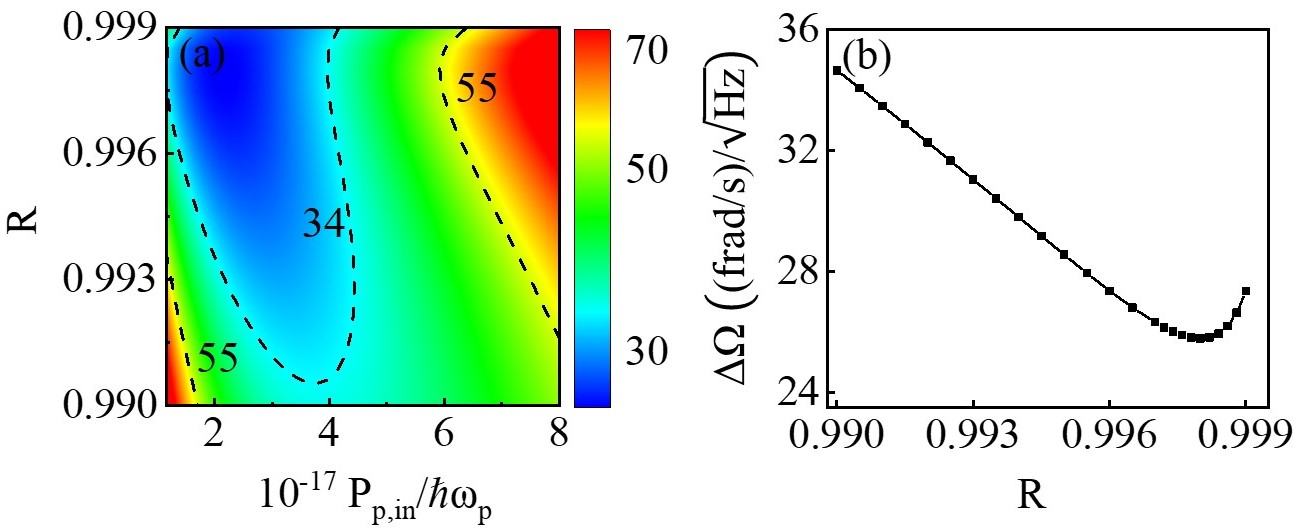} \\
	\caption{(a) Scaled sensitivity (color bar, in unit of $\femto\radian\per\second\sqrt{\hertz}$) as a function of the mirror reflectivity and the probe photon flux. (b) Highest sensitivity obtained at different mirror reflectivity when changing the probe light power.  Other parameters: $L=10~\centi\meter$, $L_0=20~\centi\meter$, $\eta = 0.5$, and $\kappa = 5$~\mega\hertz.}
	\label{fig:FIG3}
\end{figure}

According to Fig.~\ref{fig:FIG1}(b), we are interested in $R = 0.99 -0.999$, which allows us to achieve the maximum $n_\text{g}$ with low control laser power. Figure~\ref{fig:FIG3}(a) shows the gyro sensitivity versus the mirror reflectivity $R$ and the probe power $P_\text{p, in}$. The gyro is most sensitive when $R \approx 0.998$ and $P_\text{p, in}/\hbar\omega_\text{p} \approx 2 \times 10^{17}~\text{photons\per\second}$, see Fig.~\ref{fig:FIG3}. Figure~\ref{fig:FIG3}(b) clearly shows the obtained optimal sensitivity of $\Delta\Omega_\text{min} \approx 26~\femto\radian\per\second\per\sqrt{\hertz}$ at $R \approx 0.998$. A higher reflectivity of the cavity mirrors not only leads to a lower cavity loss~\cite{52.Chin. Phys. B 28}, but also significantly affects $n_\text{g}$, as shown in Fig.~\ref{fig:FIG1}(b). As a tradeoff, The mirror reflectivity is optimal at $R \approx 0.998$. Note that the slow-light enhances $n_\text{g}$ and thus the photon drag effect by a factor of $10^6$~\cite{38.Science. 356}, and significantly narrows the intracavity linewidth~\cite{41.Phys. Rev. Lett. 111}. Thanks to this doublet improvement, the gyro sensitivity can reach 9 orders of magnitude higher than Earth's rotation rate, enabling precise detection of tiny variations of the Earth's rotation and its wobbling~\cite{32.Nat. Photonics 17}.

\begin{figure}
	\centering
	\includegraphics[width=1.0\linewidth]{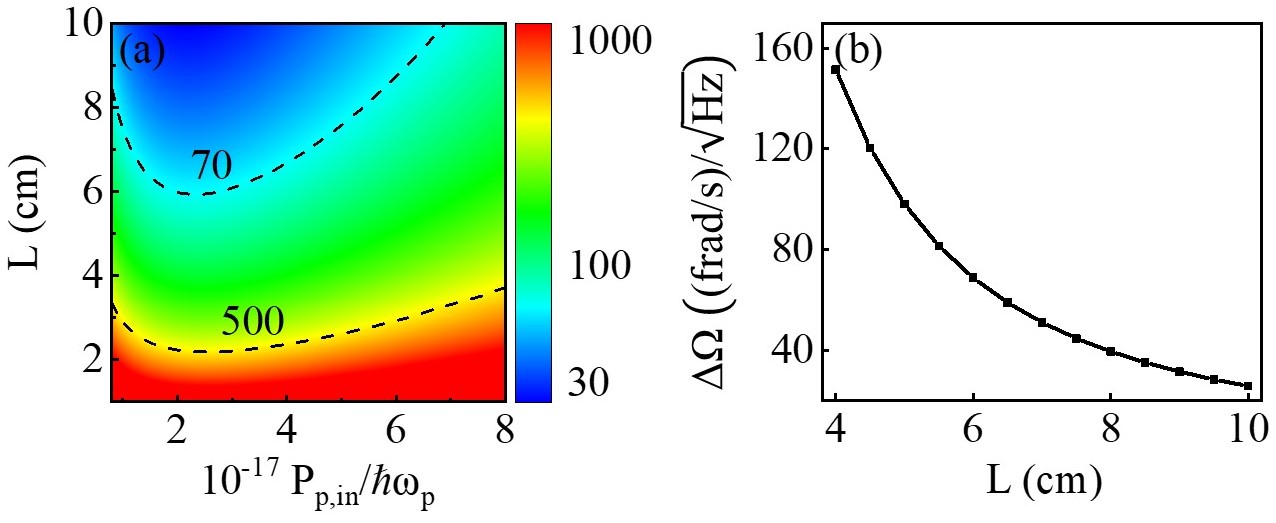} \\
	\caption{(a) Sensitivity (color bar, in unit of $\femto\radian\per\second\sqrt{\hertz}$) as a function of the medium length and the probe photon flux. (b) Optimal sensitivity obtained at different medium length by tuning the probe power. Other parameters: $R=0.998$ and $\kappa = 5$~\mega\hertz.}
	\label{fig:FIG4}
\end{figure}

Figure~\ref{fig:FIG4} presents the dependence of the gyro sensitivity on the medium length $L$. For a given $L$, the gyro sensitivity first improves as the probe photon flux increases. After reaching an optimal point, its value increases. As predicted by Eq.~\ref{eq:Delta_Omega2}, the sensitivity is inversely proportional to the duty ratio $\eta$. An example for $P_\text{p, in}/\hbar\omega_\text{p} = 2 \times 10^{17}~\text{photons\per\second} $ is presented in Fig.~\ref{fig:FIG4}(b). Obviously, a larger duty ratio yields is preferable for achieving high performance. However, considering practical experimental losses and operational, we choose a spinning medium length of $L = 10~\centi\meter$. On this basis, the gyro sensitivity can reach $26~\femto\rad /\second/\sqrt{\hertz}$. This level of sensitivity is accurate enough to detect small variations in the Earth's rotation rate, the effects of Lorentz Violation and Ggeneral Relativity in gravity research.

\section{Implementation}
In this work, we use a FP cavity embedded with a spinning medium to construct a sensitive gyro. The medium is an $\text{Er}^{3+}$-doped glass rob and slow light can be obtained with a laser beam. This laser field at wavelength $\sim 1536~\nano\meter$ is modulated in intensity with frequency $\delta$, $I=I_\text{s}+I_\text{p}\cos(\delta t)$. We fix $I_\text{p}/I_\text{s} = 0.08$, which is experimentally accessible~\cite{46.Europhys. Lett. 73}. To measure the modulated sideband light, the output VP component is measured via the homodyne detection. Specifically, the output VP light is first interfered with a local light field to generate an electric signal oscillating at frequency $\delta$. Then, this oscillating signal is filtered by an electronic circuit and measured.

%Conclusion
\section{Conclusion}\label{sec:conc}
Vibration of the Earth's rotation rate are currently measured mainly by using a large ring-laser gyroscope or fiber optic gyroscopes based on the Sagnac effect. However, these gyroscopes are large, costly, and subject to errors induced by asymmetric perturbation and common-mode noise. Here, we theoretically propose to implement a table-top gyroscope with unprecedented high sensitivity up to $26~\femto\rad /\second/\sqrt{\hertz}$ by exploring the nonreciprocity of photon drag in a rotating medium. The proposed gyroscope is small in volume and thus cost-effective. The two circularly-polarized components of laser beam co-propagate in the system, therefore eliminating the effects of the asymmetric perturbation and common-mode noise. This work opens up the door for portable gyroscope measuring the fluctuation of Earth's motion.

%Acknowledgement
\section*{Acknowledgement}
This work was supported by the National Natural Science Foundation of China (Grants No. 92365107 and No. 12305020), National Key R\&D Program of China (Grants No.~2019YFA0308700), the Program for Innovative Talents and Teams in Jiangsu (Grant No.~JSSCTD202138), the China Postdoctoral Science Foundation (Grant No.~2023M731613), and the Jiangsu Funding Program for Excellent Postdoctoral Talent (Grant No.~2023ZB708).
We thank the High Performance Computing Center of Nanjing University for allowing the numerical calculations on its blade cluster system.

\section*{APPENDIX A: COHERENT POPULATION OSCILLATORY PROCESSES IN THE $\text{Er}^{3+}$-DOPED GLASS ROD}\label{sec:APPENDIX A}

Based on the CPO in a two-level system~\cite{44.Phys. Rev. Lett. 90,45. Phys. Rev. A 24}, we can derive the density matrix equation for the $\text{Er}^{3+}$-doped glass rod. $\text{Er}^{3+}$ can be excited by a laser beam with wavelength $\lambda \approx 1536~\nano\meter$ from the ground state ${}^4I_{15/2}$ to the substable state ${}^4 I_{13/2} $ as a two-level system~\cite{42.J. Opt. Soc. Am. B 28,43.Opt. Commun 279}. Due to the long lifetime of metastable level ${}^4 I_{13/2} $ of $\text{Er}^{3+}$, the system exhibits a capacity for rapid nonlinear response within a narrow spectrum. When the control and probe fields with a small frequency difference are input simultaneously, the medium exhibits phenomena of saturable absorption.-In this situation, the number of ground state particles undergoes periodic oscillations, resulting in spectral hole burning in the absorption spectrum. At the burning point, the refractive index of light rapidly change. According to the Kramers-Kronig relationship, the group velocity also undergoes drastic changes.

For analyzing the slow light, we denote the ground state level as $\left| g \right\rangle $ and the metastable level as $\left| e \right\rangle $. According to the calculation steps in Ref.~\cite{45. Phys. Rev. A 24} , we can write the motion equation of system in the density matrix as
\setcounter{equation}{0}
\begin{subequations}\label{seq:1}
	\begin{align}
		\dot{\rho}_\text{eg}=&-\left(i\omega_\text{eg}+\frac{1}{T_2}\right)\rho_\text{eg} + \frac{i}{\hbar}V_\text{eg}\left(\rho_\text{ee}-\rho_\text{gg}\right), \\
		\begin{split}
			\dot{\rho}_\text{ee}-\dot{\rho}_\text{gg}=&\left(-\frac{1}{T_1}\right)\left[\left(\rho_\text{ee}-\rho_\text{gg}\right)-\left(\rho_\text{ee}-\rho_\text{gg}\right)^0\right]+\frac{2i}{\hbar}\\
			&\left(V_\text{ge}\rho_\text{eg}-V_\text{eg}\rho_\text{ge}\right),
		\end{split}
	\end{align}
\end{subequations}
where $T_1$ is the relaxation time of the excited state, and $T_2$ is the decoherence time of off-diagonal elements in the density matrix; $\left(\rho_\text{ee}-\rho_\text{gg}\right)^0$ is the inversion of particle number in thermal equilibrium state. $V_\text{eg}=V_\text{ge}^*=-\mu_\text{eg}\left(E_\text{s} e^{-i\omega_\text{s} t}+E_\text{p} e^{-i\omega_\text{p} t}\right)$ is the interaction Hamiltonian, where $E_\text{s}$ and $E_\text{p}$ are the amplitudes of the control and probe light, $\omega_\text{s}$ and $\omega_\text{p}$ are the frequencies of the control light and probe light, respectively, $\mu_\text{ba}$ is the dipole matrix element. When $1/T_2$ is larger than both the detuning frequency $\left(\delta=\omega_\text{p}-\omega_\text{s}\right)$ and the Rabi frequency of the control laser $\left(\Omega_\text{s} = 2 |\mu_\text{eg}| |E_\text{s}| /\hbar \right)$, the diagonal elements of the density matrix can be expressed as
\begin{equation}\label{seq:2}
	\rho_\text{eg}\approx-\frac{i\mu_\text{eg}}{\hbar} \frac{\left(E_\text{s} e^{-i\omega_\text{s} t}+E_\text{p} e^{-i\omega_\text{p} t}\right)}{i\left(\omega_\text{eg}-\omega_\text{s}\right)+1/T_2}\left[\rho_\text{ee}\left(t\right)-\rho_\text{gg}\left(t\right)\right] \;.
\end{equation}

Following the Ref.~\cite{45. Phys. Rev. A 24}, the overall particle number inversion is given by
\begin{equation}\label{seq:3}
	\rho_\text{ee}-\rho_\text{gg}\approx \left(\rho_\text{ee}-\rho_\text{gg}\right)^{dc}+\left(\rho_\text{ee}-\rho_\text{gg}\right)^\delta e^{i\delta t}+\left(\rho_\text{ee}-\rho_\text{gg}\right)^{-\delta}e^{-i\delta t},
\end{equation}
where $\left(\rho_\text{ee}-\rho_\text{gg}\right)^{dc}$ is the direct current component of the overall inversion, $\left(\rho_\text{ee}-\rho_\text{gg}\right)^{\pm \delta}$ are the amplitude components oscillating at $\delta$. When $\delta \leq 1/T_1$ , particle number oscillations are evident. 
We consider the resonance case: $\omega_\text{eg}=\omega_\text{s}$. The response at the probe light frequency can be expressed as:
\begin{equation}\label{seq:4}
	\begin{split}
		\rho_\text{eg}\left(\omega_p\right) = \frac{i\mu_\text{eg}T_2 E_\text{p}}{\hbar} \times \left[\frac{1}{1+ \Omega_\text{s}^2 T_1 T_2}- \Omega_\text{s}^2 \frac{T_2}{T_1}\frac{1+i\delta/\beta}{\delta^2+\beta^2}\right] \;,
	\end{split}
\end{equation}
where $\beta ={1}/{T_1}\left(1+ \Omega_\text{s}^2 T_1 T_2\right)$. In the two-level system, the linear susceptibility is given by $ \chi\left(\delta\right)=N\mu_\text{eg}\rho_\text{eg}\left(\omega_\text{p}\right)/E_\text{p}$. The refractive index and absorption at the hole burning location are respectively given by
\begin{equation}\label{seq:5}
	n\left(\delta\right)=1+\frac{\alpha_0 c T_1}{2\omega_\text{c}}\frac{I/I_\text{sat}}{1+I/I_\text{sat}}\left(\frac{\delta}{\left(T_1\delta\right)^2+\left(1+I/I_\text{sat}\right)^2}\right),
\end{equation}
\begin{equation}\label{seq:6}
	\alpha\left(\delta\right)=\frac{\alpha_0}{1+I/I_\text{sat}} \left(1-\frac{I/I_\text{sat}\left(1+I/I_\text{sat}\right)}{\left(T_1\delta^2\right)+\left(1+I/I_\text{sat}\right)^2}\right) ,
\end{equation}
wwhere $I$ is the intensity of input power, $I_\text{sat}$ is the saturated absorption intensity, $\alpha_0$ being the unsaturated absorption coefficient. Using the formula $n_\text{g} = n(\delta) + \omega_\text{c} dn/d\delta$, the group refractive index $n_\text{g}$ can be calculated as~\cite{44.Phys. Rev. Lett. 90}
\begin{equation}\label{seq:7}
	n_\text{g}=n_0+ \frac{\alpha_0 c T_1}{2}\frac{I/I_\text{sat}}{1+I/I_\text{sat}}\left(\frac{1}{\left(T_1\delta\right)^2+\left(1+I/I_\text{sat}\right)^2}\right),
\end{equation}
where $n_0$ is the refractive index of the host medium at $\delta = 0$ when only the control light passes through.
%
%%%%%%%%%%%%%%%%%%%%%%%%%%%%%%%%%%%%%%%%%%%%%%%%%%%%%%%%%
%\bibliographystyle{apsrev4-1}
%\bibliography{Reference}

\providecommand{\noopsort}[1]{}\providecommand{\singleletter}[1]{#1}%

\end{document}